\documentclass[prd,nofootinbib,
amsmath,amssymb,aps,
reprint,
floatfix
]{revtex4-2}

\usepackage[colorlinks=true
,urlcolor=blue
,anchorcolor=blue
,citecolor=blue
,filecolor=blue
,linkcolor=blue
,menucolor=blue
,linktocpage=true
,pdfproducer=medialab
,pdfa=true
]{hyperref}

\usepackage{hyperref}
\usepackage[T1]{fontenc}
\usepackage{graphicx}
\usepackage{comment}
\usepackage{bm}
\usepackage{braket}
\usepackage{cancel}
\usepackage{lmodern}
\usepackage{ulem}
\usepackage {diagbox}
\newcommand\aap{A\&A}                % Astronomy and Astrophysics
\newcommand\apjl{ApJ}                % Astrophysical Journal, Letters
\newcommand\jcap{J.~Cosmology Astropart. Phys.} % Journal of Cosmology and Astroparticle Physics
\newcommand\mnras{MNRAS}             % Monthly Notices of the Royal Astronomical Society
\newcommand\physrep{Phys.~Rep.}      % Physics Reports
\newcommand{\bae}[1]{\begin{align} #1 \end{align}}
\newcommand{\eq}[1]{\begin{equation}\begin{split} #1 \end{split}\end{equation}}

\newcommand{\lr}[1]{\left( #1 \right)}

\newcommand{\m}{\mathrm}

\begin{document}

\title{Cosmological free-free emission from dark matter halos in the $\Lambda$CDM model}
\author{Katsuya T. Abe}
\email{abe_kt@nagoya-u.jp}
\author{Hiroyuki Tashiro}
\affiliation{Division of Particle and Astrophysical Science,
Graduate School of Science, Nagoya University,
Chikusa, Nagoya 464-8602, Japan}

\begin{abstract}
We study the diffuse background free-free emission induced by dark matter halos. 
Since dark matter halos host ionized thermal plasma, they could be an essential source of cosmological free-free emission.
We evaluate the global background intensity and the anisotropy of this free-free emission.
We show that the dominant contribution comes from dark matter halos with a mass close to the Jeans
mass, $M_{\m{halo}}\sim 10^{10} M_\odot$, around the redshift $z \sim 3$.
Therefore, the intensity of the free-free emission is sensitive to the small-scale 
primordial curvature perturbations that form such small-mass dark matter halos.
Although our obtained intensity of the global and anisotropic free-free emission is smaller than the $10\%$ level of the free-free emission observed in the high galactic region,
we find that the free-free emission signal is modified by $\sim 20 \%$ even in the parameter set of the spectral index and the running, which is consistent with the recent Planck result. 
Therefore, the measurement of the cosmological free-free signals has the potential to provide more stringent constraints on  the abundance of small-mass dark matter halos and the curvature perturbations including the spectral index and the running,
while carefully removing the Galactic free-free emission is required through the multifrequency radio observation or the cross-correlation study with the galaxy surveys or 21-cm intensity map.
\end{abstract}

\maketitle 

\section{Introduction}\label{sec: introduction}

The {\it Planck} space mission has been achieved to measure the cosmic microwave background~(CMB) anisotropies with surprising accuracy.
The \textit{Planck} observation result is in favor of the primordial curvature perturbations which are almost scale-invariant, adiabatic, and Gaussian~\cite{2020A&A...641A...6P}.
Besides, combining the galaxy surveys~\cite{2010MNRAS.404...60R,2019JCAP...07..017C},
one can confirm that these statistical features arise from the present horizon scale to the 1~Mpc scale~\cite{2019MNRAS.489.2247C}.
Probing the statistical nature of the primordial curvature perturbations
provides the access to the early Universe model generating the primordial perturbations,
%make it possible for us to investigate the early Universe, 
especially the inflation model.
Surprisingly, obtained statistical features are consistent with the prediction from a simple inflation model, which is the slow roll inflation model with a single scalar field~\cite{2016ASSP...45...41M}~(for a review, see e.g. Ref.~\cite{2000cils.book.....L}).

Based on this great success, one of the next goals in modern cosmology is 
to reveal the primordial curvature perturbations on a smaller scale than the $\m{Mpc}$ scale.
In this context, CMB distortion can be a powerful probe~\cite{2012MNRAS.419.1294C,2014PTEP.2014fB107T,2014MNRAS.438.2065C}.
Although the small-scale perturbations 
are smoothed out due to the Silk damping, 
the energy flows dissipated in the process of the Silk damping create the distortions from the blackbody spectrum in the CMB energy spectrum~\cite{1991ApJ...371...14D,1994ApJ...430L...5H,2012MNRAS.425.1129C}.
Hence, measurements of the CMB distortion allow us to understand the small-scale perturbations more~\cite{1994ApJ...430L...5H,
2012ApJ...758...76C,2012PhRvL.109b1302P, 2012PhRvD..86b3514D,2013MNRAS.434.1619C}.
In fact, from the measurements of CMB distortion by COBE/FIRAS,
the constraint of the primordial power spectrum, $P_\zeta \lesssim 10^{-5}$ for the wave number range, $k \approx 1-10^4~\rm Mpc^{-1}$ was suggested in Ref.~\cite{2012ApJ...758...76C}.
Furthermore, it is suggested that in the next-generation CMB measurements like PIXIE
\cite{2019arXiv190901593C}, the constraint would improve in order of 
$P_\zeta \lesssim 10^{-8}$.

Future measurements of the redshifted 21-cm signal are also expected to be helpful in measuring the small-scale perturbations.
Since the 21-cm signal is emitted by the transition of the neutral hydrogen, in which the electron flips its spin,
the measurements of the spatial fluctuations in the redshifted 21-cm signal can trace the evolutionary history of
the matter density fluctuations
%of neutral hydrogen 
before the epoch of reionization~(for a comprehensive review, see~Ref.\cite{2006PhR...433..181F}).
Redshifted 21-cm measurements have the potential to explore the density perturbations on smaller scales than the Silk scale~\cite{2004PhRvL..92u1301L}.
To accomplish these measurements, the Square Kilometre Array project, building the largest radio telescope with over a square kilometer of collecting area, is proceeding now.

Observations of the early-formed minihalo signature are also suggested in cosmological observations
to constrain the small-scale density fluctuations.
If the fluctuations on smaller scales have 100 times larger amplitude than the scale-invariant spectrum confirmed by Planck, a large amount of minihalos forms after the matter-radiation equality~\cite{2009ApJ...707..979R}. 
Because of such early formation, their core density is higher than one formed in the standard hierarchical structure formation. 
The measurement of the $\gamma$-ray sky by Fermi can provide a tight constraint 
on the abundance of minihalos and small-scale perturbations if dark matter~(DM) can self-annihilate~\cite{2009PhRvL.103u1301S,2010PhRvD..82h3527J,2012PhRvD..85l5027B}.
Redshifted 21-cm emission has also been investigated as a probe of these early formed minihalos, which does not require the self-annihilating DM model~\cite{2002ApJ...572L.123I,2014JCAP...03..001S,2014PhRvD..90h3003S,
2018JCAP...02..053S,2020MNRAS.494.4334F}.
In addition, recently, the free-free emission from these minihalos has been studied, and using the Planck free-free measurement, the author provides the tighter constraint on the small-scale structure~\cite{2022PhRvD.105f3531A}.

In this work, we focus on the diffuse background free-free emission as the probe of the small-scale perturbation.
The diffuse background free-free emission has been studied in the context of the foreground components e.g. the CMB in the microwave and radio frequency range.
Although most of the observed free-free emission is considered to be composed of Galactic origin,
cosmological free-free emission is generated, for example, in the intergalactic medium~(IGM)~\cite{2004ApJ...606L...5C},
the galaxy groups and clusters after the reionization~\cite{2011MNRAS.410.2353P},
and the structure formation during the reionization~\cite{2019MNRAS.486.3617L}.
Among them, the contribution from DM halos could be the largest~\cite{2008MNRAS.391..383G}.

Here, we revisit the free-free emission from DM halos in the standard $\Lambda$CDM cosmology from the point of view of the small-scale primordial curvature perturbation. In particular, we investigate the dependency on the spectral index $n_{\m{s}}$ of the primordial power spectrum and the running $r_{n_{\m{s}}}$ which appears as the cosmological parameters for the first time.
To investigate it, we first study the distributions of redshift and DM halo mass for both the global signal and the anisotropy of the free-free emission. After that, we prepare several parameter sets $(n_{\m{s}},r_{n_{\m{s}}})$ which would modify the number density evolution, especially for small-mass halos, and test how the emission signal and the anisotropy change by the different parameter sets. We also discuss the dependence on the gas profile model within DM halos.

This paper is organized as follows.
In Sec.\ref{sec: ff_from_ind_halos}, we provide
the halo model describing the profile of their gas density and temperature.
Then, we calculate the intensity of the free-free emission from an individual halo for different mass $M_\m{halo}$.
In Sec.~\ref{sec: bgff_from_lcdmhalos},
considering the halo formation history,
we formulate the diffuse background intensity
which is the sum of the free-free emission
from individual halos.
We also discuss the mass and redshift distribution of the disuse background intensity.
In Sec.~\ref{sec: anisotropy_ff_from_LCDMhalos}, we formulate the anisotropies of the free-free emission.
We also estimate the mass and redshift distribution of the anisotropies in the same way as the one of the diffuse background intensity.
In Sec.~\ref{sec: spectral_index}, we discuss the application of our results to the constraint on the primordial curvature perturbations.
We conclude in Sec.~\ref{sec: conclusion}.

In this paper, we apply the flat $\Lambda$CDM cosmology and use the best-fit cosmological parameters of the latest Planck result~\cite{2020A&A...641A...6P}.
We also use the calculation package named HMFcalc~\cite{murray2013hmfcalc} to estimate the halo mass function, the matter power spectrum, and so on.

\section{free-free emission rate from an individual halo}\label{sec: ff_from_ind_halos}

Thermal plasma can emit free-free radiation. 
When the number density, temperature, and ionization fraction 
of the free electrons in the plasma
are given by $n_{\rm gas}$, $T_{\rm gas}$ and $x_{\rm e}$,
the emission coefficient of the free-free radiation 
at a frequency~$\nu$ is given 
by~\cite{Radiative_process_in_Astrophysics}
\begin{align}\label{eq: brems_emit_rate}
\epsilon_{\nu}^{\m{ff}}
=& \frac{2^3  e^6}{3 m_\m{e} c^3}
\left(\frac{2 \pi}{3 m_\m{e} k_{\rm B} T_\m{gas}}\right)^{1/2} \nonumber \\
&\times x_{\rm e}^2{ n^2_{\rm gas}}
\exp({- h_{\m{p}} \nu/k_{\rm B} T_\m{gas}})
\bar{g}_{\m{ff}}~,
\end{align}
where $e$ and $m_e$ is the electric charge and mass of electrons, $h_{\m{p}}$ is the Planck constant, and 
$k_{\m B}$ is the Boltzman constant.
In Eq.~\eqref{eq: brems_emit_rate},
$\bar{g}_{\m{f f}}$ is a {velocity-averaged Gaunt factor}.
We adopt the fitting formula of $\bar{g}_{\m{f f}}$ in Ref.~\cite{2011piim.book.....D},
\eq{\label{eq: gaunt_factor}
\bar{g}_{\m{f f}}=\log \left\{\exp \left[5.960-\sqrt{3} / \pi \log \left(\nu_{9}T_{4}^{-3 / 2}\right)\right]+\mathrm{e}\right\},
}
where $\nu_{9}\equiv \nu/(1~\m{GHz})$,
$T_{4} \equiv T_\m{gas}/(10^4~{\m K})$,
and $\m{e}$ is Napier's constant.
The free-free emission at a frequency larger than the critical frequency, $\nu_{\rm c} = {k_{\rm B} T_\m{gas}/h_{\m{p}}}$,
suffers the exponential damping as shown in Eq.~\eqref{eq: brems_emit_rate}.
However, since we are interested in the CMB or radio frequency range, this damping effect can be almost negligible.
% Since these frequency range is well below the critical frequency
% given by the gas temperature in DM halos,
% we assume $\exp(- h_{\m{p}} \nu/k_{\rm B} T_\m{halo}) \approx 1$ hereafter.

To evaluate the free-free emission from DM halos,
it is required to model the gas profile for the energy density and temperature in DM halos.
In this paper, we adopt the gas profile~(KS model) given in Refs.~\cite{Komatsu_2001,2002MNRAS.336.1256K}.

In the KS model, the gas density and temperature profiles are written by
\bae{
\rho_{\mathrm{gas}}(x)&=\rho_{\mathrm{gas}}(0) y_{\mathrm{gas}}(x),
\\
T_{\mathrm{gas}}(x)&=T_{\mathrm{gas}}(0) y_{\mathrm{gas}}^{\gamma-1}(x),
}
where $x$ is given by $x=r/R_{\rm vir}$ with the virial radius $R_{\m{vir}}$,
$\rho_{\mathrm{gas}}(0)$, and $T_{\mathrm{gas}}(0)$ are the number density and temperature at $x=0$,
$\gamma$ is the polytropic index,
and $y_{\mathrm{gas}}$ is the shape function of the profile which satisfies
$y_{\rm gas} (0)=1$.
Here the shape function, $y_{\mathrm{gas}}$, is found by solving the hydrostatic equilibrium equation.
In this work, we employ the DM NFW density profile~\cite{NFWprofile1997} following Ref.~\cite{Komatsu_2001} although actual halos-hosted galaxies may have a different shallower profile (see e.g. Ref.~\cite{2019A&ARv..27....2S} for the review.).
Then one can obtain
\eq{
y_{\mathrm{gas}}\equiv\left\{1-B\left[1-\frac{\ln (1+x)}{x}\right]\right\}^{1 /(\gamma-1)},
}
with
\eq{
B \equiv 3 \eta^{-1}_0 \frac{\gamma-1}{\gamma}\left[\frac{\ln (1+c_s)}{c_s}-\frac{1}{1+c_s}\right]^{-1},
}
where $\eta_0$ is the mass-temperature normalization factor which is defined by
\eq{
\eta_0\equiv  \frac{3k_{\rm B} R_{\m{vir}} T_{\rm gas}(0)} {G\mu m_{\rm p}  M_{\m{halo}}},
\label{eq:eta0}
}
and $c_s$ is the concentration parameter.
In Eq.~\eqref{eq:eta0}, $G$ is the gravitational constant, $\mu$ represents the mean molecular weight (we set $\mu = 0.6$), and $m_{\m{p}}$ is the proton mass. 
For the values of $\eta_0$ and $\gamma$, Ref.~\cite{2002MNRAS.336.1256K} provides the useful fitting formula as functions of $c_s$,
\eq{\label{eq: def_eta0}
\eta_0 \approx 2.235+0.202(c_s-5)-1.16 \times 10^{-3}(c_s-5)^{2},
}
and
\eq{\label{eq: def_gamma}
\gamma=1.137+8.94 \times 10^{-2} \ln (c_s / 5)-3.68 \times 10^{-3}(c_s-5).
}
These functions are valid in range of $0<c_s<25$.

Reference~\cite{2020arXiv200714720I} has investigated the concentration parameter~$c_s$ by performing 
N-body simulations from the high redshift to the present.
Based on the simulation results,
they provide the analytic fitting formulae of $c_s$
for the DM halos with mass $M_\m{halo}\gtrsim 10^9 ~M_\odot$ in the redshift range of $0< z < 14$.
In this paper, we employ their result for the concentration parameter $c_s$.
We confirmed that $c_s$ is less than $c_s<25$
for the DM halo mass range and the redshifts investigated in this paper. Therefore, the fitting formulae of Eqs.~\eqref{eq: def_eta0} and \eqref{eq: def_gamma} must be valid throughout the paper.

With $\eta_0$ in Eq.~\eqref{eq:eta0}, we can estimate $T_{\m{gas}}(0)$ as
\eq{\label{eq: def_Tgas0}
T_{\mathrm{gas}}(0)
=&2\eta_{0}
\left(\cfrac{\mu}{0.6}\right)
\lr{\cfrac{M_\m{halo}}{10^{10} h^{-1} M_{\odot}}}\lr{\frac{R_{\m{vir}}}{67 h^{-1}\m{kpc}}}^{-1}\m{[eV]}
.}
It is noted that the Eq.~\eqref{eq: def_Tgas0} with $\eta_0=1$ shows a value of the virial temperature of the halo with a mass $M_{\m{halo}}$, $T_{\m{vir}}(M_{\m{halo}})$. 
The integration of the density profile provides 
$\rho_{\rm{gas}}(0)$ as
\eq{\label{eq: def_rhogas0}
\rho_{\rm{gas}}(0)&={M_{\rm{gas}}}
\left[
{4 \pi r_{\rm s}^{3} \int^{u_{\rm v}}_{0} y_{\rm{gas}}(u)u^2du}
\right]^{-1}\\
&\ = 2.63\times 10^{12} \mathrm{M}_{\odot} \mathrm{Mpc}^{-3} \\ & \times\left(\frac{\Omega_{\mathrm{b}} h^{2}}{\Omega_{\mathrm{m}}}\right) \lr{\cfrac{M_\m{halo}}{10^{10} h^{-1} M_{\odot}}}\lr{\frac{R_{\m{vir}}}{67 h^{-1}\m{kpc}}}^{-3} \\ & \times c_s^{3}\left[\frac{y_{\mathrm{gas}}^{-1}(c_s)}{c_s^{2}(1+c_s)^{2}}\right]\left[\ln (1+c_s)-\frac{c_s}{1+c_s}\right]^{-1},
}
where $M_{\m{gas}}$ is the total baryonic mass contained in the halo with a mass, $M_\m{halo}$.
Here we assume that the dark matter halos can host the baryon gas whose mass is given by
\eq{\label{eq: Mgas_from_Mhalo}
M_{\m{gas}} = \frac{\Omega_{\m{b}}}{\Omega_{\m{m}}} M_{\m{halo}}.
}

The ionization fraction is determined by the balance between the recombination and the ionization by thermal collision or UV photons from galaxies and stars.
Most DM halos contribute to the free-free radiation form after the Epoch of reionization. 
Therefore, we simply assume that UV photons from galaxies and stars are enough to keep the ionization in DM halos, that is we set $x_{\m{e}}=1$.

\section{Diffuse background free-free emission induce by DM halos}\label{sec: bgff_from_lcdmhalos}

The free-free emission from an individual DM halo can be evaluated by Eq.~\eqref{eq: brems_emit_rate} with the gas profile discussed in the previous section. 
Now we calculate the global intensity of the diffuse free-free background emission which is the sum of the emission from DM halos in the Universe.

To calculate the global intensity, it is useful to evaluate the
mean intensity of the free-free emission from an individual 
halo with a mass $M_{\m{halo}}$ at a redshift~$z$,
\begin{align}
\label{eq: Inu_w/_impact_param}
I_{\nu}^{\m{ind}}(z,M_\m{halo}) = 
\frac{\int_{V_{\rm halo}} \epsilon_{\nu}^{\m{ff}}dV}{ S_{\m{halo}}},
\end{align}
where ${V_{\rm halo}}$ and $S_{\m{halo}}$ is the physical volume and 
cross section on the sky for the DM halo,
%$V_{\m{halo}}$ is the physical cross section of the DM halo on the sky,
$V_{\m{halo}} = 4 \pi R_{\m{vir}}^3/3$ and $S_{\m{halo}} = \pi R_{\m{vir}}^2$,
respectively.
Note that, to obtain Eq.~\eqref{eq: Inu_w/_impact_param},
we apply the optically thin approximation, because 
the free-free absorption is negligible in our case.

Let us consider the redshift shell
at a redshift $z$ with the width~$dz$.
The free-free emission contribution from DM halos in this redshift shell is 
\eq{\label{eq: dI_sky_dz}
&dI_\m{\nu}^{\m{ff}}(z)
=
dz \frac{dV_{\rm com}}{dz} 
\int_{M_{\min}} dM_\m{halo}~
%f_{\rm sky} 
\frac{\Omega_{\m{halo}}}{4\pi} I_{\nu}^{\m{ind}}
\frac{dn_{\m{halo}}^{\m{com}}}{{~d} M_{\m{halo}}},
}
where
$dn_{\rm halo}^{\rm com} /dM_{\m{halo}}$ is the comoving mass function of DM halos, $V_{\rm com}$ is the comoving volume, 
%\eq{\label{eq: sky_z}
%f_{\rm sky } = %\frac{\Omega_{\m{halo}}}{4\pi},
%}
and $\Omega_{\m{halo}}(z,z_{\rm f}, M_\m{halo})$ is the solid angle of a DM halo given by $\Omega_{\m{halo}} = \pi{((1+z)R_{\m{vir}})^2}/{\chi^2}$
with the comoving distance $\chi$ to a redshift $z$.
In Eq.~(\ref{eq: dI_sky_dz}), 
we set the DM minimum mass $M_{\min}$ to the Jeans mass 
with the background baryon temperature $T_{\rm b,bg}$.
Since we are interested in the redshifts after the epoch of reionization, we take
$T_{\rm b,bg} \sim 10^4~\rm K$~\cite{2000MNRAS.318..817S,2011MNRAS.410.1096B,2012MNRAS.424.1723G,2014MNRAS.438.2499B,2014MNRAS.441.1916B}.
Below $M_{\min}$, the baryon gas cannot collapse at the formation of DM halos,
and DM halos can retain baryon gas through accretion, which is a relatively smaller amount than through collapsing. 
Therefore, we neglect the contribution from such small-mass DM halos.

Finally, the redshift integration yields 
the total global
intensity at an observed frequency $\nu_{\rm obs}$ from DM halos,
\eq{\label{eq: Inu_full}
I_{\m{obs}} (\nu_{\m{obs}})
=\int^{\infty}_{0}~dz~\frac{1}{(1+z)^3}
\frac{dI_\m{\nu_{\m{em}}}^{\m{ff}}(z,M_\m{halo})}{dz},
}
where
$\nu_{\rm em} $ is $\nu_{\rm em} = (1+z)\nu_{\rm obs} $ and 
the factor $(1+z)^{-3}$ is the redshift effect for the intensity.

\begin{figure}[btp]
    \centering
    \includegraphics[width=1.0\hsize]{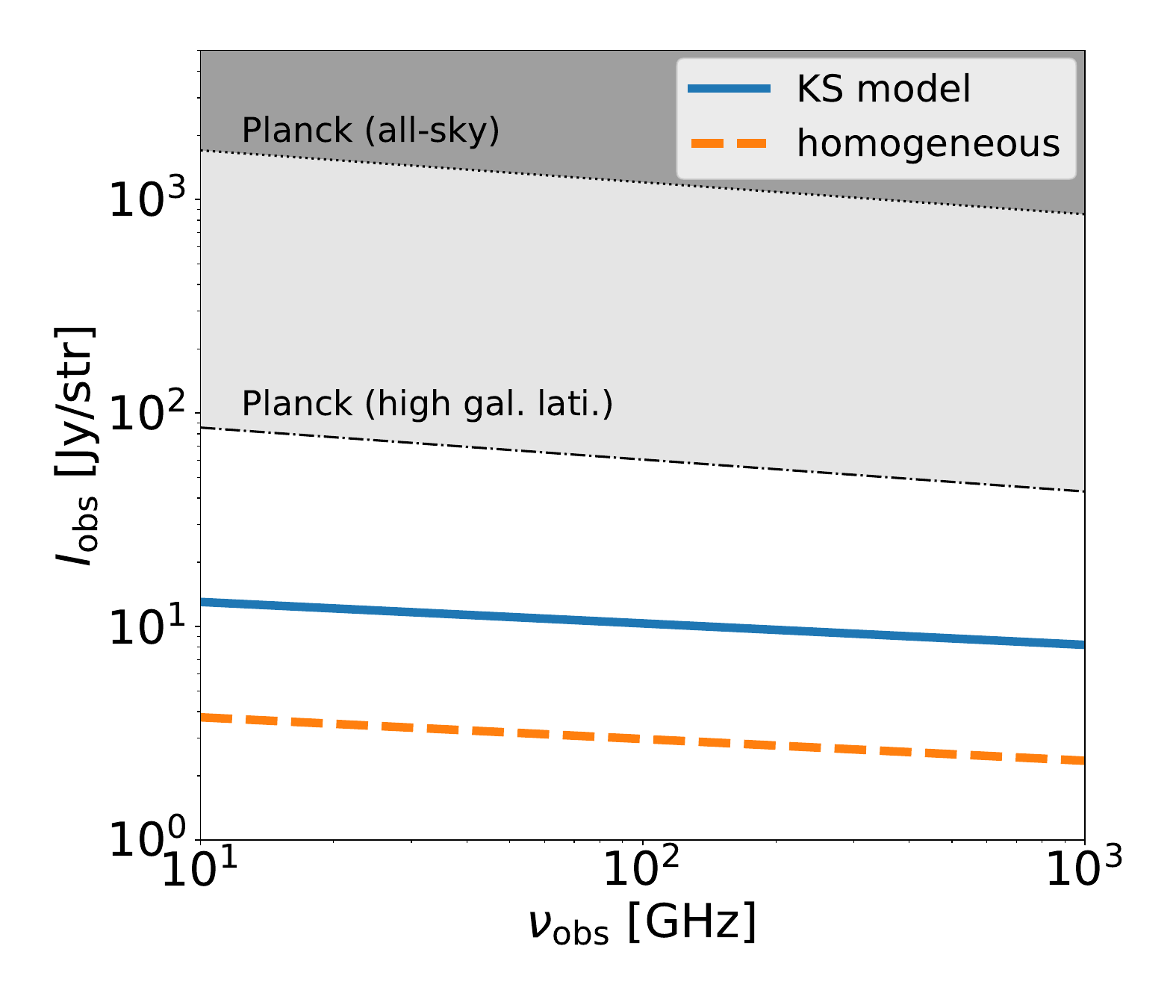}
    \caption{The intensity of the global free-free emission from DM halos. The horizontal axis is for the frequency in a unit of $\m{GHz}$. The blue solid line shows the global intensity for the KS model, and the orange dashed line is the one for the homogeneous gas model estimated by Eq.~\eqref{eq: Inu_full}. The dark(light) gray shaded region is excluded by the all-sky(high galactic latitude) mean intensity observed by Planck.
    }
    \label{fig: Iobs_twocases}
\end{figure}

We plot the result of Eq.~\eqref{eq: Inu_full} in Fig.~\ref{fig: Iobs_twocases}.
To obtain this result, we adopt the Press-Schechter mass function with the Planck best-fit cosmological parameter set.
It might be helpful to represent the intensity in terms of the brightness temperature,~$T_{\m{b},\nu}$.
In the Rayleigh-Jeans limit,
the brightness temperature is related to the intensity by
\eq{
T_{\m{b},\nu}=\frac{c^2}{2\nu_{\m{obs}}^2}I_{\m{obs}}\simeq 0.32\lr{\frac{I_{\m{obs}}}{1~\m{Jy/str}}}\lr{\frac{\nu_{\m{obs}}}{10~\m{GHz}}}^{-2}[\mu\m{K}].
}
Therefore, while the intensity of the free-free signals is almost frequency independent,
the brightness temperature of the free-free signal is proportional to $\nu^{-2}$.

Since the free-free intensity is proportional to the square of the gas number density,
the strength of the intensity highly depends on the gas profile model in DM halos, for which we take the KS model explained in Sec.~\ref{sec: ff_from_ind_halos}.
In order to show the impact of the gas profile, we also calculate the intensity in the homogeneous gas model where $n_{\m{gas}}=200n_{\m{gas,bg}}$ with the background IGM gas number density $n_{\m{gas,IGM}}$ and $T_{\m{halo}}=T_{\m{vir}}(M_{\m{halo}})$, and plot the intensity as an orange dashed line in Fig.~\ref{fig: Iobs_twocases}.

\begin{figure}[btp]
    \centering
    \includegraphics[width=1.0\hsize]{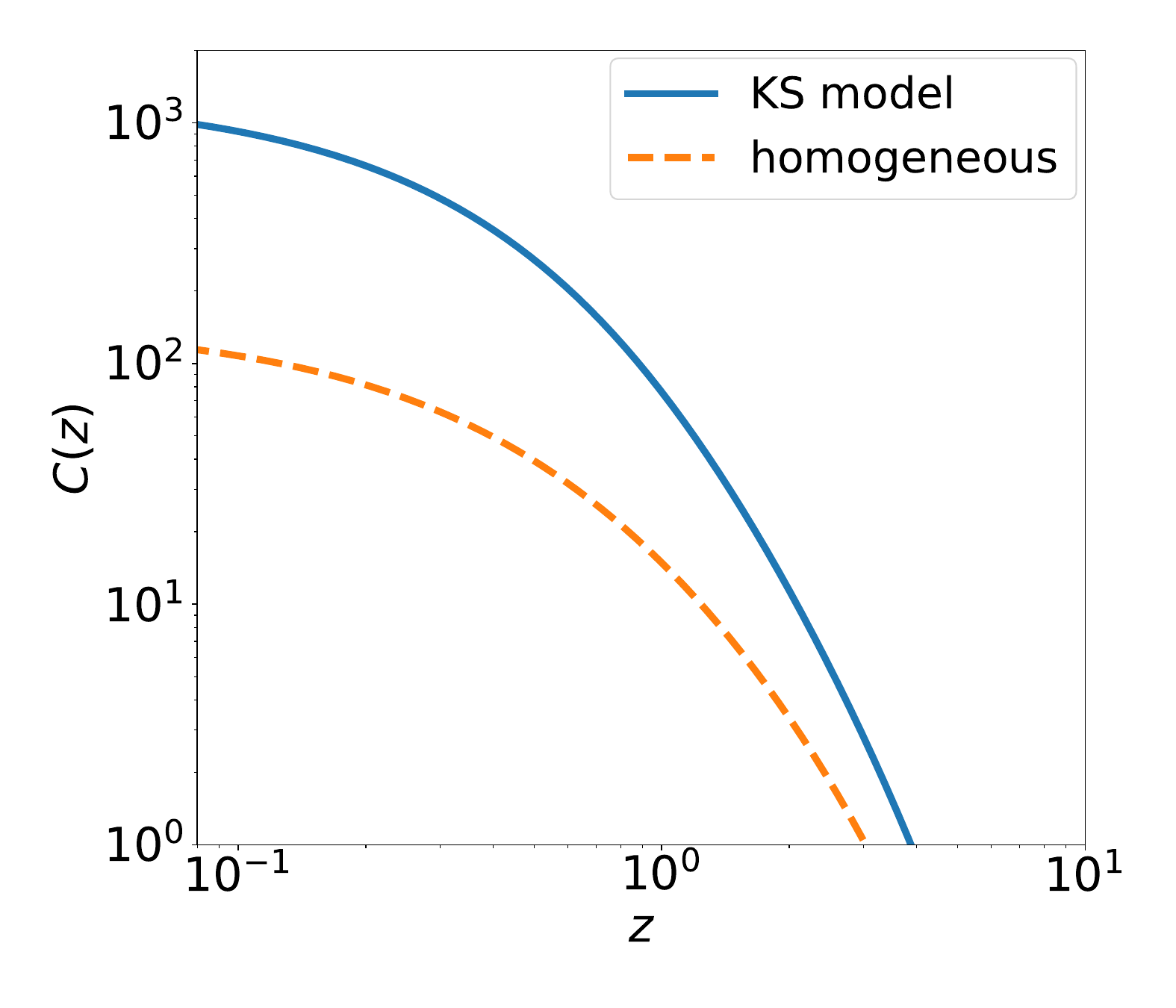}
    \caption{The redshift dependence of the clumping factor in Eq.~\eqref{eq: clump}. The blue solid and orange dashed lines are for the KS model and the homogeneous gas model, respectively.}
    \label{fig: clumping_factor}
\end{figure}

The clumpiness is a good indicator representing the impact of the gas profile model on the free-free emission and is defined as
\eq{\label{eq: clump}
C(z)\equiv \frac{\int dM_{\m{halo}}\int_{V_{\m{halo}}} dV n_{\m{e}}^2(r,M_{\m{halo}})\frac{dn_{\m{halo}}^{\m{com}}}{dM_{\m{halo}}}}{\bar{n}_{\m{e},\m{bg}}^2},
}
where $n_{\m{e}}(r,M_{\m{halo}})=x_{\rm e}n_{\rm gas}$ is the radial electron number density in an individual halo, and $\bar{n}_{\m{e},\m{bg}}^2$ represents the mean value of the free electron number density in the IGM.
We show the clumpiness of both models in Fig.~\ref{fig: clumping_factor}.
The clumpiness in the KS model is roughly 10 times larger than in
the homogeneous model. However, the free-free intensity in the KS model is
not amplified as expected from the clumping factor. This is because the gas temperature in DM halos also has the profile in the KS model, and it weakens the amplification from the enhancement of the clumpiness as you can see in Eq.~\eqref{eq: brems_emit_rate}.
Note that the homogeneous model could provide the minimum intensity of the free-free emission from DM halos, compared with other gas models.
Therefore, Fig.~\ref{fig: Iobs_twocases}
tells us that the difference in the gas profile modifies the intensity
by at least a factor of a few.

The free-free emission in the CMB frequency has been studied well as the foreground of the CMB.
In Fig.~\ref{fig: Iobs_twocases},
we plot the (all-sky) mean free-free intensity provided by the Planck collaboration and the one at the high galactic latitude given in Ref.~\cite{2022PhRvD.105f3531A}.
The current observed signal even at high galactic latitude is 10 times larger than our prediction with the Planck best-fit parameters.
In the observed free-free signals,
the cosmological contribution has not been identified yet,
although most of the observed free-free signals are considered to be of Galactic origin.
To identify the cosmological contribution as we predict here,
further investigation including removing Galactic signals, analyzing the statistical anisotropy of the signal, and taking the cross-correlation with cosmological observations are required.

\subsection{Mass and redshift distribution}

The free-free signals calculated above are the sum of the contribution from all DM halos spread in the wide range of the mass and redshift.
It is worth investigating the mass and redshift contribution of the global intensity.

\begin{figure}[tbp]
    \centering
    \includegraphics[width=1.0\hsize]{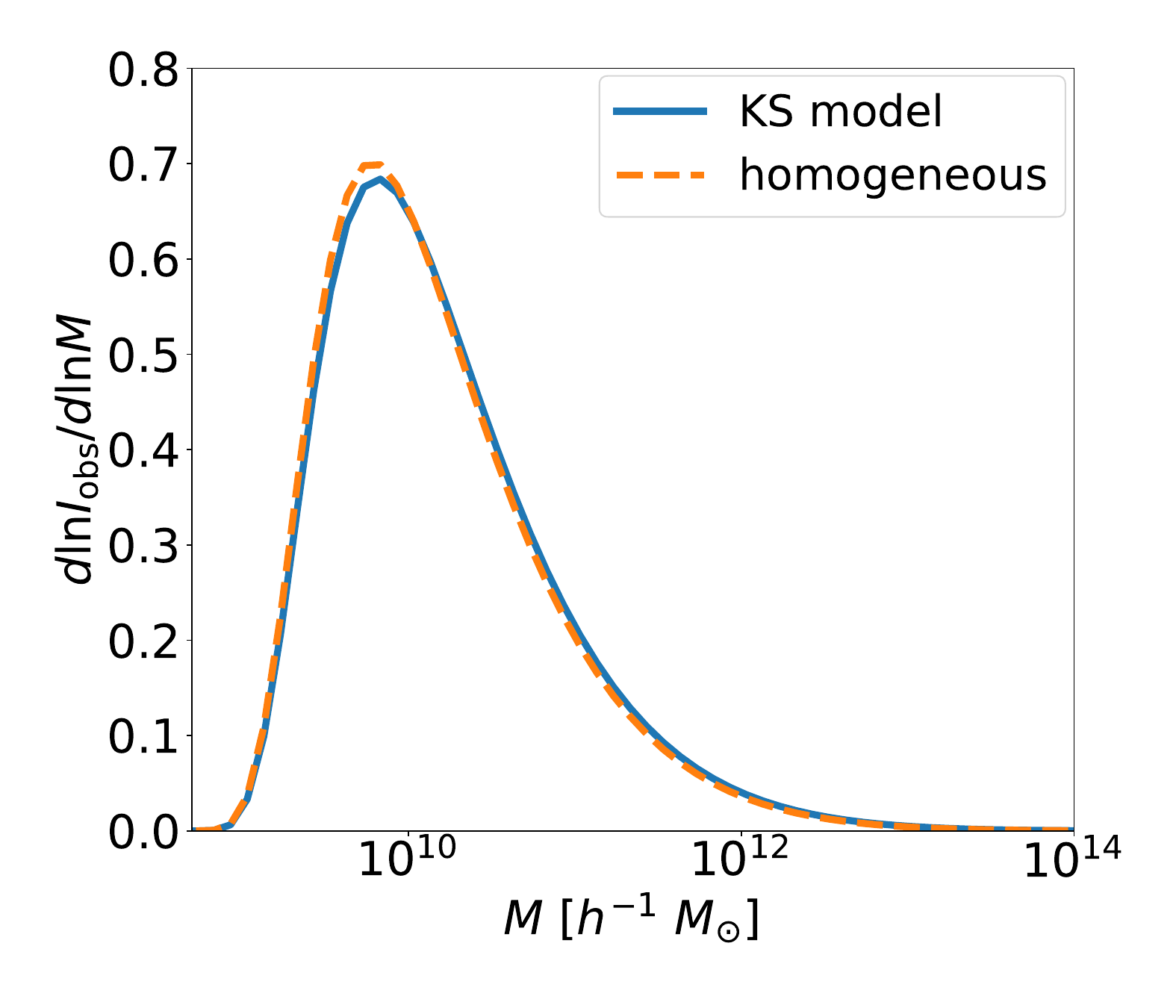}
    \caption{Mass distribution of the global free-free emission from DM halos.
    The blue solid and orange dashed lines are for the KS model and the homogeneous gas model, respectively.}
    \label{fig: dIobsdlnMdz}
\end{figure}

The halo mass distribution of $I_{\m{obs}}$ at $\nu_{\m{obs}}=70~\m{GHz}$ is represented in Fig.~\ref{fig: dIobsdlnMdz}.
The figure clearly shows that
the dominant contribution comes from relatively small-mass halos around $M_{\m{halo}}\sim10^{10}M_{\odot}$.
The monotonic decrease in a large mass side is
due to the shape of the halo mass function.
As the DM mass increases,  
the number density of DM halos becomes small, and as a result, the intensity contribution decreases.
The cutoff on the lower mass side is due to the Jeans mass.
Therefore, most of the free-free emission signals come from DM 
halos around the Jeans mass scale.
The concentration parameter also depends on the mass.
As DM halos have a smaller mass, the concentration parameter would increase, and the resultant signal would be enhanced.
However, we found that the mass distributions of the signals for both the KS and the homogeneous gas models are almost identical.
This is because the contribution coming from the halo mass function is much larger than the one of the concentration parameter.

We also plot the redshift distribution for these two models in the same way as Fig.~\ref{fig: dIobsdlnMdz}.
Basically, the redshift distribution reflects the history of the formation efficiency of DM halos, especially with $M_{\m{peak}}\sim 10^{10}M_{\odot}$ which is corresponding to the peak in Fig.~\ref{fig: dIobsdlnMdz}. 
That is, the peak feature represents in the redshift $z_{\m{peak}}\sim 3$ when $\sigma(M_{\m{peak}},z_{\m{peak}})/\delta_{c}\sim 1$. Unlike the mass distribution, the evolution of the concentration parameter significantly contributes to this redshift distribution.
As shown in Fig.10 of Ref.~\cite{2020arXiv200714720I}, for the halo with a mass around $M_{\m{halo}}\sim M_{\m{peak}}$, the concentration parameter increases as the redshift becomes small.
Therefore, the redshift distribution for the KS model shifts to the smaller redshift side than in the homogeneous model.

Because of this redshift dependence, we suggest that 
if we obtained the redshift tomographic information on the global signals of free-free emission through e.g., the cross-correlation analysis with 21-cm line intensity, we could know about the small-scale structure formation in the Universe and the gas profile in small-mass halos.

\begin{figure}[hbtp]
    \centering
    \includegraphics[width=1.0\hsize]{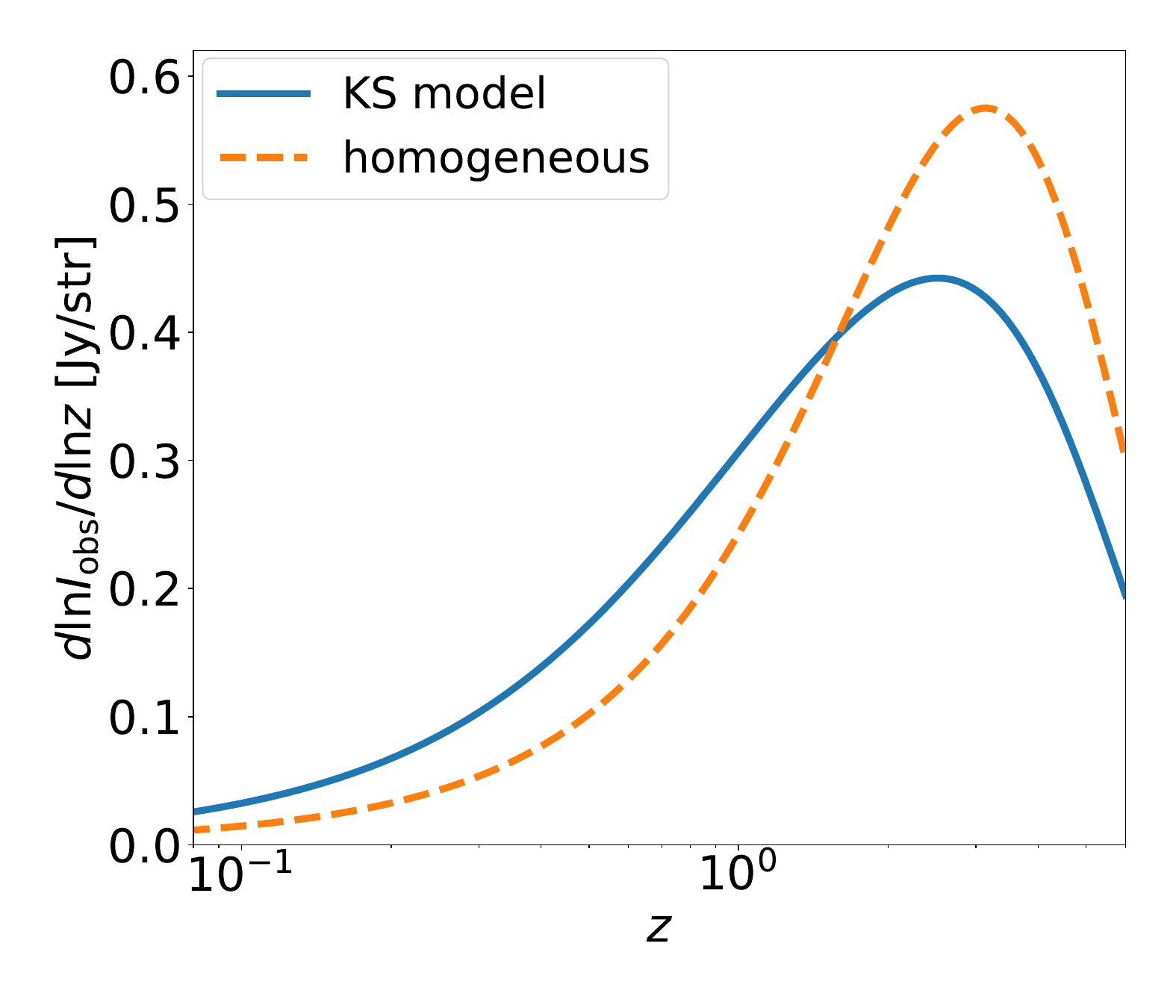}
    \caption{Redshift distribution of the global free-free emission from DM halos.
     The blue solid line is for the KS model, and the orange dashed line is for the homogeneous gas model.
    }
    \label{fig: dIobsdz}
\end{figure}

\section{Statistical Anisotropy of the free-free emission from DM halos}\label{sec: anisotropy_ff_from_LCDMhalos}

In this section, we investigate the anisotropy of the free-free emission induced by DM halos.
The anisotropy of the cosmological free-free signals is created by
both the clustering of DM halos and the Poisson contribution to the number density of DM halos.
The statistical value of the anisotropy is evaluated in terms of the angular power spectrum.
To compute the angular power spectrum, we adopt the halo formalism of Ref.~\cite{Halo_formalism_cooray}.
Accordingly, the power spectrum can be divided into two components,
\eq{\label{cleq} 
C_{\ell}^{\m{ff}}=C_{\ell}^{1 {\m{h}}}+C_{\ell}^{2{\m{h}}}.
}
Here $C_{\ell}^{1{\m{h}}}$ is the ``one-halo'' term describing the Poisson contribution, and $C_{\ell}^{2{\rm h}}$ is the ``two-halo'' term describing the clustering contribution. With the limber approximation, they can be written by~\cite{1980lssu.book.....P,Cole&Kaiser_Cly}
% Here $C_{\ell}^{1{\rm h}}$ is the ``one-halo'' term describing the Poisson contribution written by
\eq{\label{eq: Cl_1halo_term}
C_{\ell}^{1{\rm h}}=\int_{0}^{\infty}
dz~\frac{d^2V}{dzd\Omega}\int dM_{\m{halo}}~\frac{dn_{\rm{halo}}^{\m{com}}}{dM_{\m{halo}}}\left|\tilde{I}_{\ell}(z)\right|^{2},
}
% and $C_{\ell}^{2{\rm h}}$ is the ``two-halo'' term describing the clustering contribution. 
and
% $C_{\ell}^{2{\rm h}}$ can be written by
\eq{\label{eq: Cl_2halo_term}
C_{\ell}^{2\rm{h}}\approx \int^{\infty}_{0}
 dz\frac{d^2V}{dzd\Omega}P\left(\frac{\ell}{\chi(z)},z \right)
 \left|\int dM_{\m{halo}}~ \tilde{\Psi}(M_{\m{halo}},z)\right|^{2},
}
where $P\left(k,z\right)$ is the conventional matter power spectrum at the redshift $z$.
In Eq.~\eqref{eq: Cl_2halo_term}, $\tilde{\Psi}$ is defined as
\eq{
\tilde{\Psi}\equiv \frac{dn_{\rm{halo}}^{\m{com}}}{dM_{\m{halo}}}\tilde{I}_{\ell}(M_{\m{halo}},z)b(M_{\m{halo}},z),
}
where $\tilde{I}_{\ell}$ is the 2D Fourier modes of the intensity, and $b$ is the halo bias which we employ in Ref.~\cite{Halo_bias_Mo_White}.

Since our gas model is spherically symmetric, the 2D Fourier modes of the intensity are calculated by
\eq{
\tilde{I}_{\ell} = \frac{1}{d_{\m{A}}^2}\int dR R^2 \epsilon_{\nu}^{\m{ff}}(R) \frac{\sin(\ell R/d_{\m{A}})}{\ell R/d_{\m{A}}},
}
where $R$ is the proper radial distance from the center of the halo, and $d_{\m{A}}=d_{\m{A}}(z)$ is the proper angular  diameter distance.

\begin{figure}[htbp]
    \centering
    \includegraphics[width=1.0\hsize]{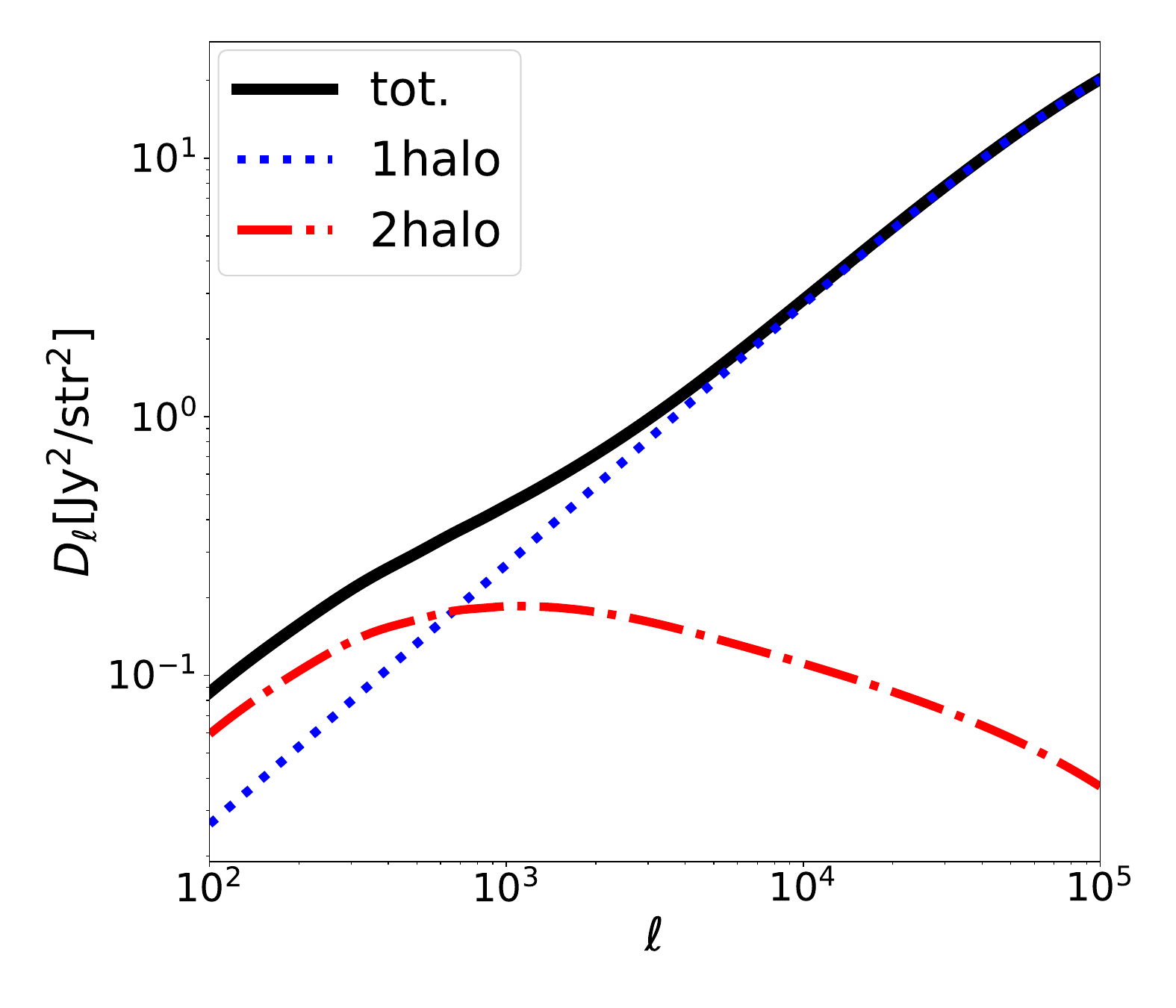}
    \caption{Angular power spectrum of the cosmological free-free emission induced by DM halos. The blue dotted line shows the one-halo term in Eq.~\eqref{eq: Cl_1halo_term}, the red dash-dot line represents the two-halo term in Eq.~\eqref{eq: Cl_2halo_term}, and the black solid line is the total.}
    \label{fig: D_ell_II_ff}
\end{figure}

We calculate the angular power spectrum of the free-free emission in the KS model
and show the result, $D_\ell \approx \ell^2/2\pi C_{\ell} ^{\rm ff}$, in Fig.~\ref{fig: D_ell_II_ff}.
We set the frequency to $\nu_{\m{obs}}=10\m{[GHz]}$ for the calculation of the angular power spectrum. 
The blue dotted line shows the one-halo term contribution in Eq.~\eqref{eq: Cl_1halo_term}, and the red dashed-dotted line represents the two-halo term one in Eq.~\eqref{eq: Cl_2halo_term}. The total angular spectrum is represented by the black solid line in Fig.~\ref{fig: D_ell_II_ff}.
Figure~\ref{fig: D_ell_II_ff} tells us that,
on large scale,
the one- and two-halo terms create the contribution at the same level.
On the other hand, 
on small scales~($\ell\gtrsim 10^3$), the dominant contribution of the anisotropy comes from 
the one-halo term rather than the two-halo term.
We also mention the observational aspect. The CMB frequency range $~\mathcal{O}(10-100)\mathrm{GHz}$ is a sweet spot for the observation of not only CMB but also free-free emissions. Thus to detect the free-free emission anisotropy from the standard dark matter halos, one should be able to observe fluctuations at least a level of order $1\m{Jy/str}$ on angular scales of $0.1$ deg to $0.01$ deg in this frequency range.

\subsection{Mass and redshift distribution}

\begin{figure}[htp]
    \centering
    \includegraphics[width=1.0\hsize]{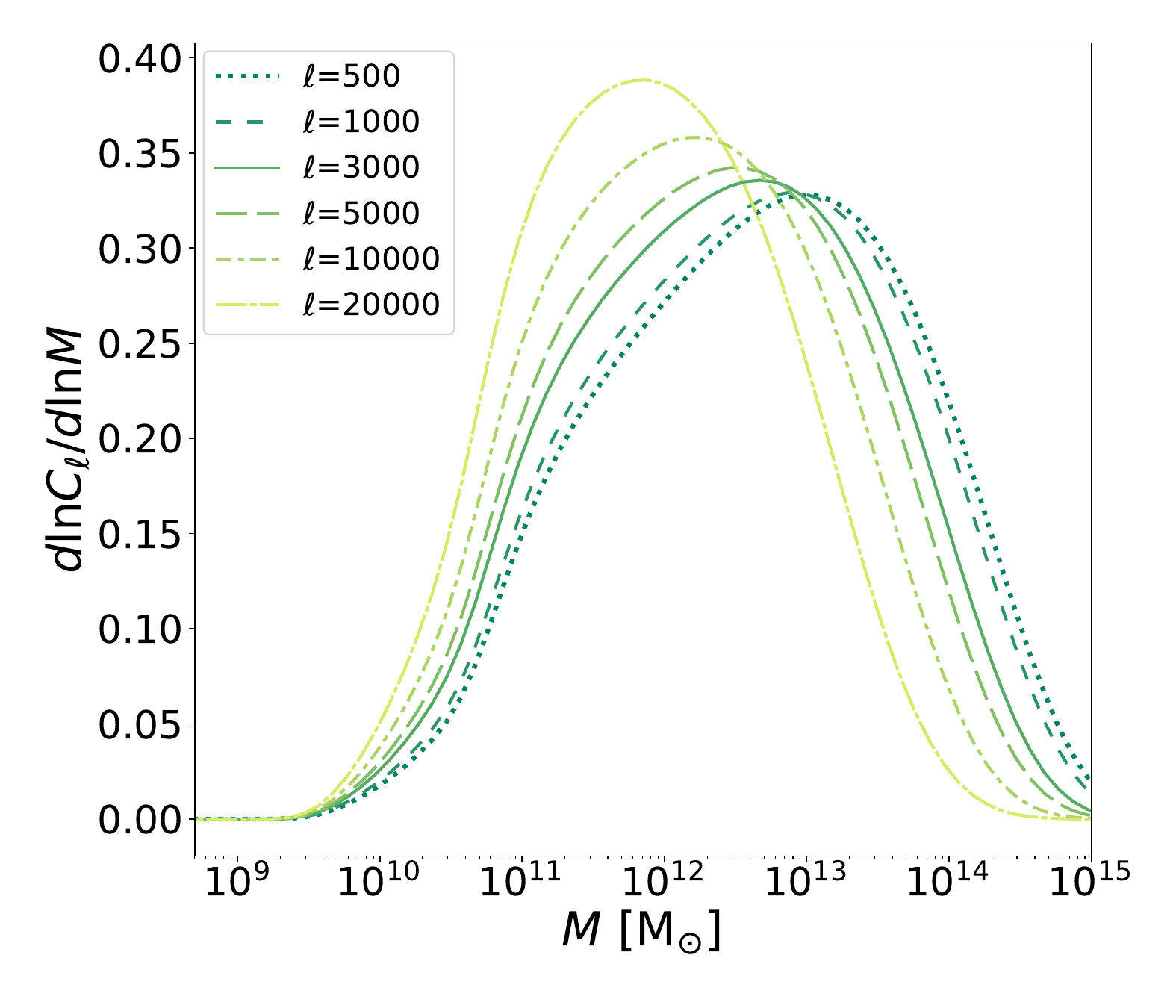}
    \caption{Mass distribution of the one-halo term of the free-free emission anisotropy induced by the distribution of DM halos in Eq.~\eqref{eq: dlnC_ell_dlnM_1h}.
    The color light and shade correspond to the smaller and larger angular scale respectively.}
    \label{fig: dlnC_ell_dlnM_1h}
\end{figure}

Here we investigate the mass and redshift distribution in the anisotropy of the cosmological
free-free emission.
First, we consider the mass distribution of the one-halo term.
This mass distribution can be evaluated by
\eq{\label{eq: dlnC_ell_dlnM_1h}
\frac{d \ln C_{\ell}^{\m{1h}}}{d \ln M_{\m{halo}}} \equiv \frac{M_{\m{halo}} \int d z \frac{d V}{d z} \frac{d n^{\m{com}}_{\m{halo}}(M_{\m{halo}}, z)}{d M_{\m{halo}}}\left|\tilde{I}_{\ell}(M_{\m{halo}}, z)\right|^{2}}{\int d z \frac{d V}{d z} \int d M_{\m{halo}} \frac{d n^{\m{com}}_{\m{halo}}(M_{\m{halo}}, z)}{d M_{\m{halo}}}\left|\tilde{I}_{\ell}(M_{\m{halo}}, z)\right|^{2}}.
}
Figure~\ref{fig: dlnC_ell_dlnM_1h} shows the mass distribution for different angular scales.
As the $\ell $ mode becomes small, the peak location of the contribution shift to large-mass halos.
As discussed above, small-mass halos produce the dominant contribution.
However, in the anisotropy,
the contribution of small mass halos on small $\ell$ is suppressed proportionally to $\ell^2$
because of the Poisson contribution. As a result, the profile structure of large mass halo 
cannot be relatively neglected on large scales, compared with small-mass halo contribution.

Next, we study the redshift contribution.
The redshift distributions are obtained through
\eq{\label{eq: dlnC_ell_dlnz_1h}
\frac{d \ln C_{\ell}^{\m{1h}}}{d \ln z} \equiv \frac{z \frac{d V}{d z} \int d M_{\m{halo}} \frac{d n_{\m{halo}}^{\m{com}}(M_{\m{halo}}, z)}{d M_{\m{halo}}}\left|\tilde{I}_{\ell}(M_{\m{halo}}, z)\right|^{2}}{\int d z \frac{d V}{d z} \int d M_{\m{halo}} \frac{d n_{\m{halo}}^{\m{com}}(M_{\m{halo}}, z)}{d M_{\m{halo}}}\left|\tilde{I}_{\ell}(M_{\m{halo}}, z)\right|^{2}},
}
and
\eq{\label{eq: dlnC_ell_dlnz_2h}
\frac{d \ln C_{\ell}^{\m{2h}}}{d \ln z} \equiv \frac{z \frac{d V}{d z} P\lr{\frac{\ell}{\chi(z)}}\left|\int d M_{\m{halo}} \tilde{\Psi}(M_{\m{halo}},z)\right|^{2}}{\int d z \frac{d V}{d z} P\lr{\frac{\ell}{\chi(z)}}\left|\int d M_{\m{halo}} \tilde{\Psi}(M_{\m{halo}},z)\right|^{2}}.
}

\begin{figure}[htp]
    \centering
    \includegraphics[width=1.0\hsize]{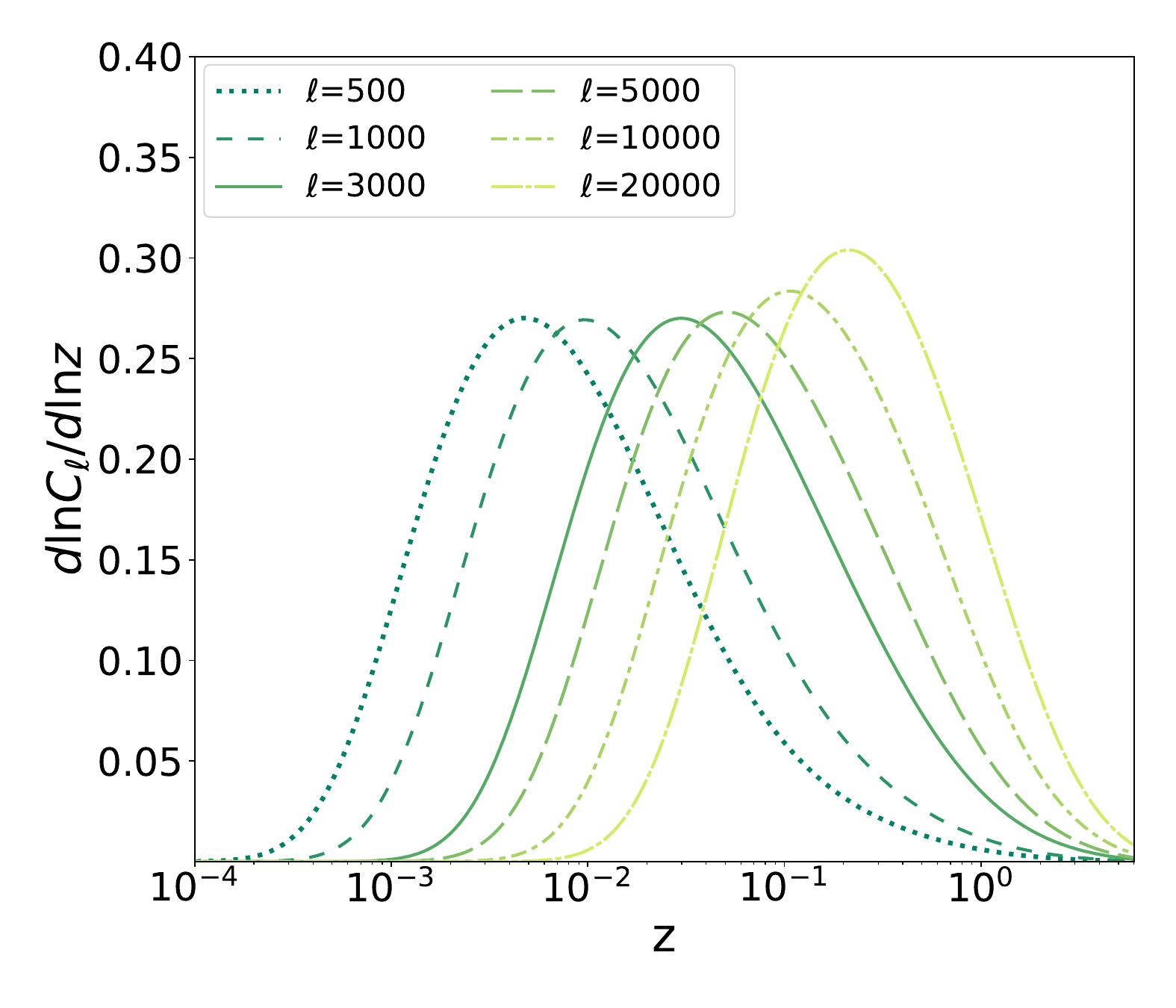}
    \caption{Redshift distribution of the one-halo term of the free-free emission anisotropy induced by the distribution of DM halos in Eq.~\eqref{eq: dlnC_ell_dlnz_1h}.
    The color light and shade correspond to the smaller and larger angular scale respectively.}
    \label{fig: dlnC_ell_dlnz_1h}
\end{figure}

\begin{figure}[htp]
    \centering
    \includegraphics[width=1.0\hsize]{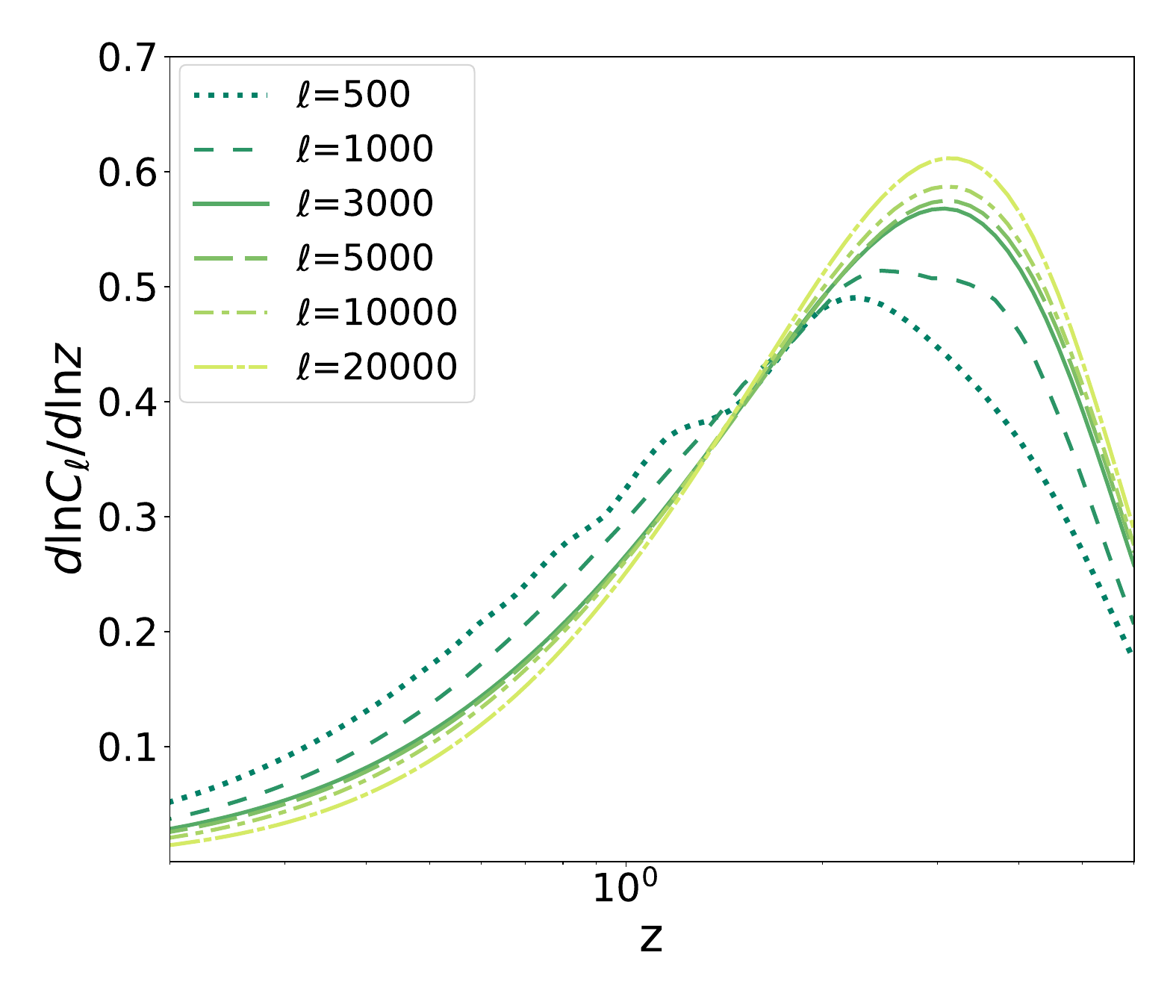}
    \caption{Redshift distribution of the two-halo term of the free-free emission anisotropy induced by the distribution of DM halos in Eq.~\eqref{eq: dlnC_ell_dlnz_1h}.
    The color light and shade correspond to the smaller and larger angular scale respectively.}
    \label{fig: dlnC_ell_dlnz_2h}
\end{figure}

Figures~\ref{fig: dlnC_ell_dlnz_1h} and~\ref{fig: dlnC_ell_dlnz_2h} show the redshift distribution for the one- and two-halo terms respectively.
The dependence of the peak location on $\ell$ modes
is different between the one- and two-halo terms.
In the one-halo term, the peak location depends on $\ell$ modes. It moves toward lower redshifts as the $\ell$ mode decreases. This is because the apparent size
of DM halos is important through $I_\ell$.
As shown in the mass contribution of the one-halo term in Fig.~\ref{fig: dlnC_ell_dlnM_1h}, most of the contributions come from small-mass halos as the Poisson contributions. On small $\ell$ modes,
the larger apparent angle size the DM halos have,
the smaller the suppression of the Poisson noise is, because 
the suppression is proportional to $\ell/\ell_{\rm halo}^3$, where 
$\ell_{\rm halo}$ is the $\ell$ mode corresponding to the apparent angular size of halos.
Therefore, since DM halos at lower redshift have large apparent angular scales, 
the peak locates on lower redshifts as the $\ell$ mode becomes small.
On the other hand, in the two-halo terms,
all $\ell$ modes have a sharp peak at the same redshift, $z\sim 3$.
The two-halo term has 
a strong dependence on the abundance of DM halos with $M_{\m{halo}} \sim 10^{10}~{M}_\odot$
because of $C_{\ell}^{\m{2h}} \propto (dn_{\m{halo}}/dM_{\m{halo}})^2$.
DM halos with $M_{\m{halo}} \sim 10^{10}~{M}_\odot$ actively form at $z\sim 3$.

\section{Cosmological application: the spectral index and running of the primordial curvature perturbations}\label{sec: spectral_index}

As discussed in the previous sections with Figs.~\ref{fig: dIobsdlnMdz} and~\ref{fig: dlnC_ell_dlnM_1h},
most contribution of the global signal and the anisotropy of free-free emission comes from small-mass DM halos whose mass is corresponding to around the Jeans scale. 
This suggests that the observations of the cosmological free-free signal might be a good tool to probe the abundance of such DM halos.
Probing the abundance of small-mass halos is important to study the statistics of the primordial curvature perturbations, in particular, on small scales.
In this section, we investigate the dependence of the free-free signals on the statistic of the primordial curvature perturbations.

The statistical nature of the primordial curvature perturbations can be provided in the nondimensional primordial curvature power spectrum $\mathcal{P}_{\zeta}$.
To describe the $k$ dependence, the spectral index, $n_s$, and the running, $r_{n_{\m{s}}}$ are often introduced as
\eq{\label{eq: power_def}
\mathcal{P}_{\zeta}&=\lr{\frac{k}{k_{\m{pivot}}}}^{n(k)},
\\
n(k)&\equiv n_{\m{s}}-1 + \frac{1}{2}
%\frac{dn_{\rm{s}}}{d \rm{ln} k}
r_{n_{\m{s}}}\ln \lr{\frac{k}{k_{\m{pivot}}}},
}
where $k_{\m{pivot}}$ is the pivot scale given in $k_{\m{pivot}}=0.05\m{Mpc}^{-1}$.

According to the latest Planck paper, the spectral index and the running are evaluated as
\eq{\label{eq: ns_run_planck}
n_{\m{s}} = 0.9659 \pm 0.0040 \quad(69\% \m{C.L.}),\\
r_{n_{\m{s}}}
%\equiv \frac{dn_{\rm{s}}}{d \rm{ln} k} 
= - 0.0041 \pm 0.0067 \quad(69\% \m{C.L.}).
}

In this work, we set six parameter combinations, $(n_{\m{s}},r_{n_{\m{s}}})$ summarized in the Table~\ref{table: ns_rns} based on Eq.~\eqref{eq: ns_run_planck} to investigate the dependency of the cosmological free-free signal 
on the spectral index.
The first three models~(columns I-III) are no-running models but with the best-fit value of the spectral index and the maximum values in the $1\sigma$- and $2\sigma$-region.
The latter three models~(columns IV-VI) have not only the tilts but also the running of the best-fit value and the $1\sigma$- and the $2\sigma$-maximum values for the running, respectively.\footnote{It is noted that in the Planck analysis, the estimated value from the cosmological model with running and the one without running are different. However, in this work, we call the value estimated by the cosmological model including the running index the best-fit value even in our model without the running.}
% Our parameter set is consistent with the CMB anisotropy measured by Planck.

\begin{table}[htp]
\begin{tabular}{|c|c|c|c|c|c|c|}
\hline
    & I & II & III & IV & V & VI\\
    \hline
$n_{\m{s}}$ & 0.9659  & 0.9699 & 0.9739 & 0.9659  & 0.9699 & 0.9739\\
\hline
% $r_{n_{\m{s}}}$ &\diagbox[height=\line]{\ }{\ } &\diagbox[height=\line]{\ }{\ }&\diagbox[height=\line]{\ }{\ }&-0.0041&0.0026& 0.0093\\ 
$r_{n_{\m{s}}}$ & 0 & 0 & 0 &-0.0041&0.0026& 0.0093\\ 
\hline
\end{tabular}
\caption{Parameter sets}
\label{table: ns_rns}
\end{table}

\begin{figure}[htp]
    \centering
    \includegraphics[width=1.0\hsize]{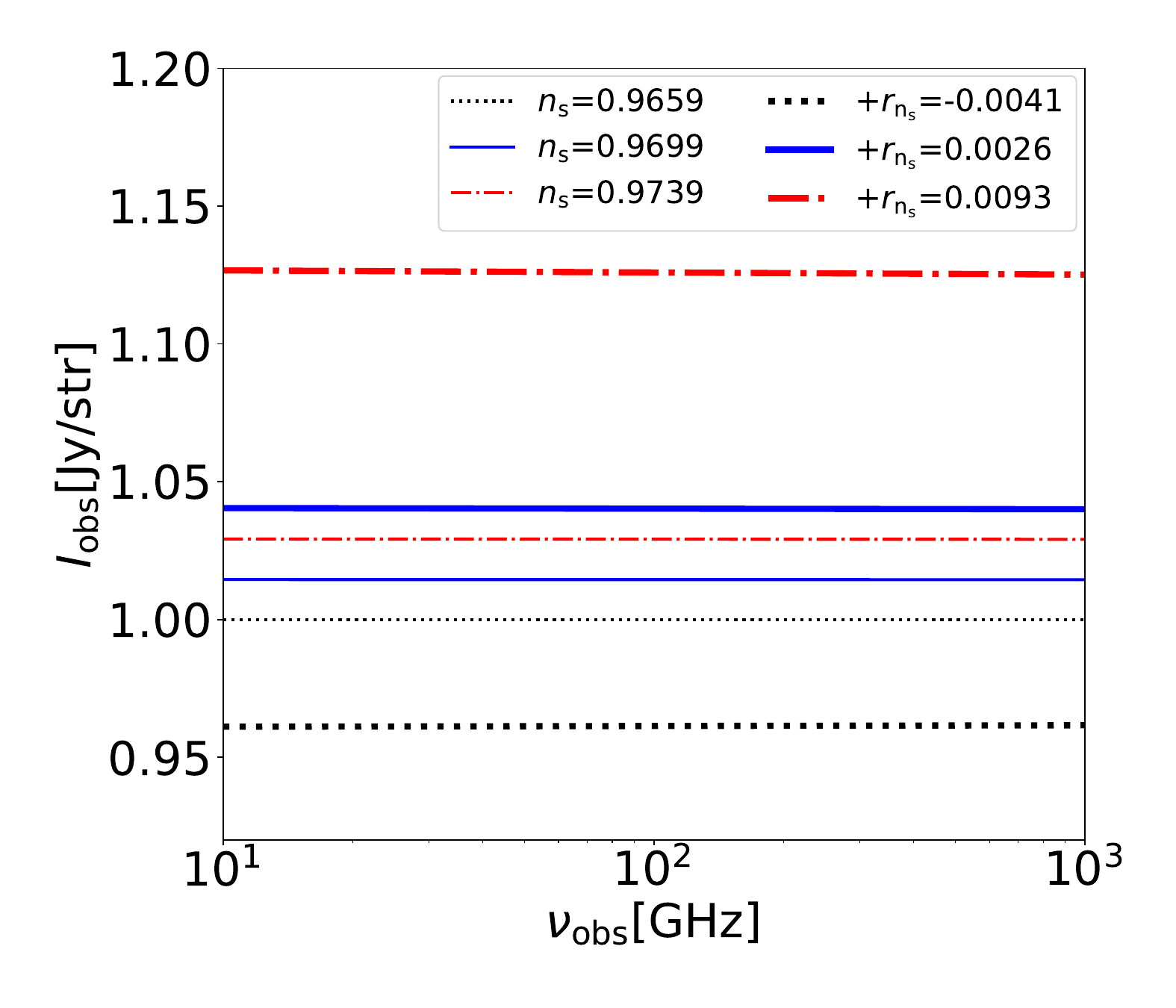}
    \caption{The global signals of halos' free-free emission with the six parameter sets are summarized in Table.~\ref{table: ns_rns}.}
    \label{fig: Iobs_ns_run}
\end{figure}

Figure~\ref{fig: Iobs_ns_run} shows the global signals with six parameter sets.
Although the frequency dependences with the six parameter sets are the same, the amplitudes are different. 
As the spectrum becomes close to blue tilt, the signal amplitude would increase. This is because the abundance of small-mass DM halos, which mainly contribute to the free-free signals, would be enhanced.
The impact of the running paramter $r_{n_{\m{s}}}$ is prominent. 
Even in the parameter set consistent with the Planck data, the amplitude of the free-free signal would be enhanced by $\sim 12\%$.

\begin{figure}[htp]
    \centering
    \includegraphics[width=1.0\hsize]{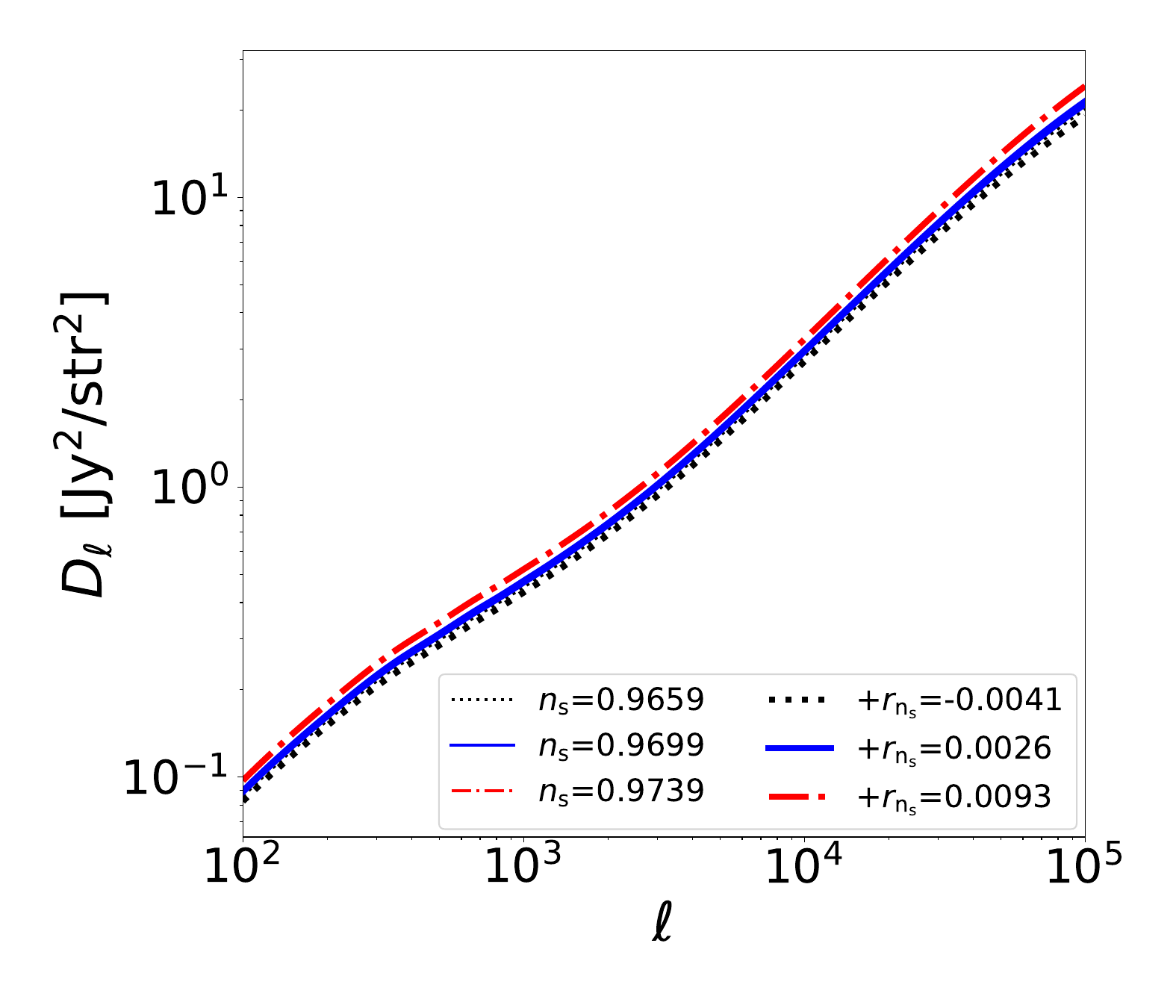}
    \caption{The anisotropies of halos' free-free emission signal with the six parameter sets are summarized in Table.~\ref{table: ns_rns}.}
    \label{fig: Dell_ns_run_pl}
\end{figure}

\begin{figure}[htp]
    \centering
    \includegraphics[width=1.0\hsize]{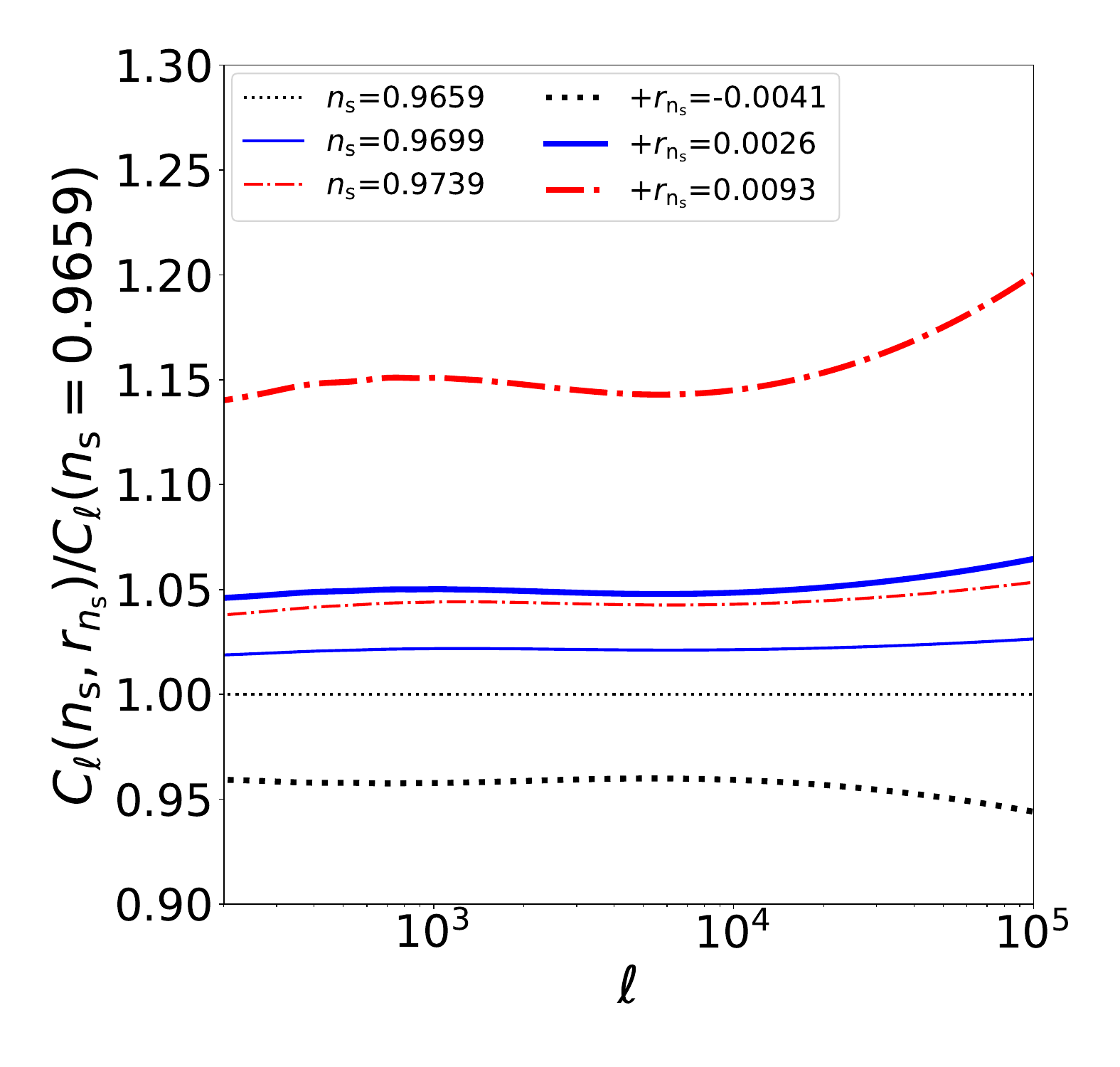}
    \caption{The difference of anisotropies of free-free emission by halos with the six parameter sets summarized in Table.~\ref{table: ns_rns}. We normalized them by the parameter set (I), $(n_{\m{s}},r_{n_{\m{s}}})=(0.9659,0)$.}
    \label{fig: Cell_ratio_ns_run_pl}
\end{figure}

%\subsection{the anisotropy}
The anisotropies of the free-free emission for the six parameter sets are plotted in Fig.~\ref{fig: Dell_ns_run_pl}.
For comparison, we provide the ratio of the anisotropy of free-free emission to the parameter set (I) for the other parameter sets in Fig.~\ref{fig: Cell_ratio_ns_run_pl}.
Similar to the global signal, the anisotropy signal is enhanced in the models which have a large amplitude of the primordial curvature perturbations on large $k$ scales.
In particular, the enhancement becomes large on small scales and reaches $\sim 20~\%$ amplification for the parameter set VI.


It is worth comparing this anisotropy dependence on the parameter set with the one of CMB signals; the CMB primordial temperature anisotropy and the thermal Sunyaev-Zel'dovich~(SZ) effect.

\begin{figure}[htp]
    \centering
    \includegraphics[width=1.0\hsize]{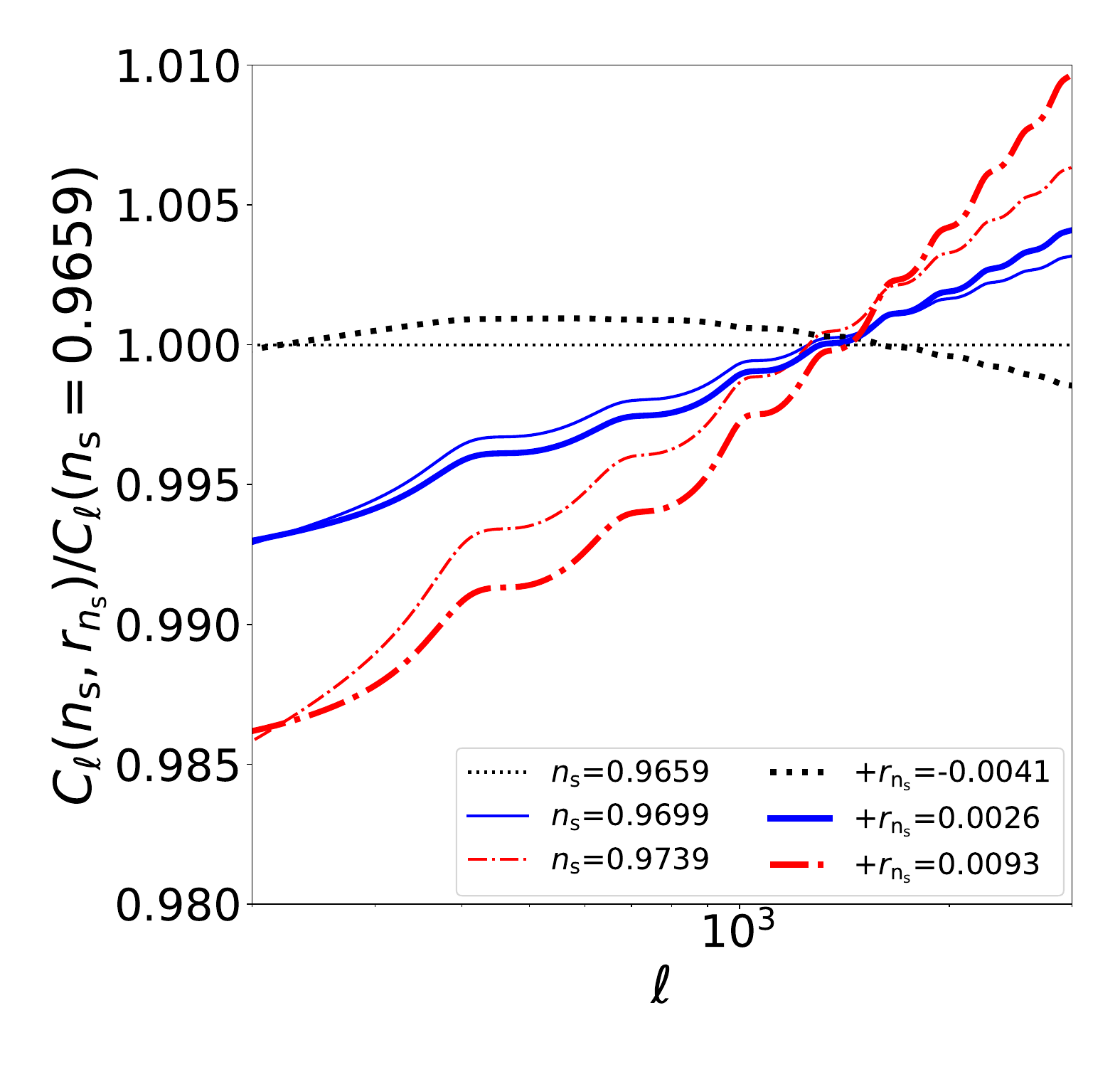}
    \caption{The difference of CMB temperature anisotropies with the six parameter sets summarized in Table.~\ref{table: ns_rns}. We normalized them by the parameter set (I), $(n_{\m{s}},r_{n_{\m{s}}})=(0.9659,0)$.}
    \label{fig: Cell_TT_ratio_ns_run_pl}
\end{figure}

The CMB primordial temperature anisotropy is one of the great probes to know the primordial curvature perturbations. 
We know its powerfulness by the Planck analysis of cosmological parameter sets mentioned above.
Similarly to Fig.~\ref{fig: Cell_ratio_ns_run_pl}, we plot the ratio of the CMB temperature anisotropy for the six parameter sets in Fig.~\ref{fig: Cell_TT_ratio_ns_run_pl}.
As the scale deviates from the pivot scale corresponding to $\ell \sim 1000$, the difference of the ratio from the unity would increase.
For example, on a small scale, $\ell \sim 3000$ the signal enhancement is, at most, $\sim 1\%$ among six parameter sets. 
It is expected that the difference becomes large on smaller scales.
However, on such smaller scales, the primordial temperature anisotropies are significantly damped by the Silk damping, and it would be challenging to measure them.
Therefore, it is difficult to probe the small-scale primordial curvature perturbations through the primordial temperature perturbations.

\begin{figure}[htp]
    \centering
    \includegraphics[width=1.0\hsize]{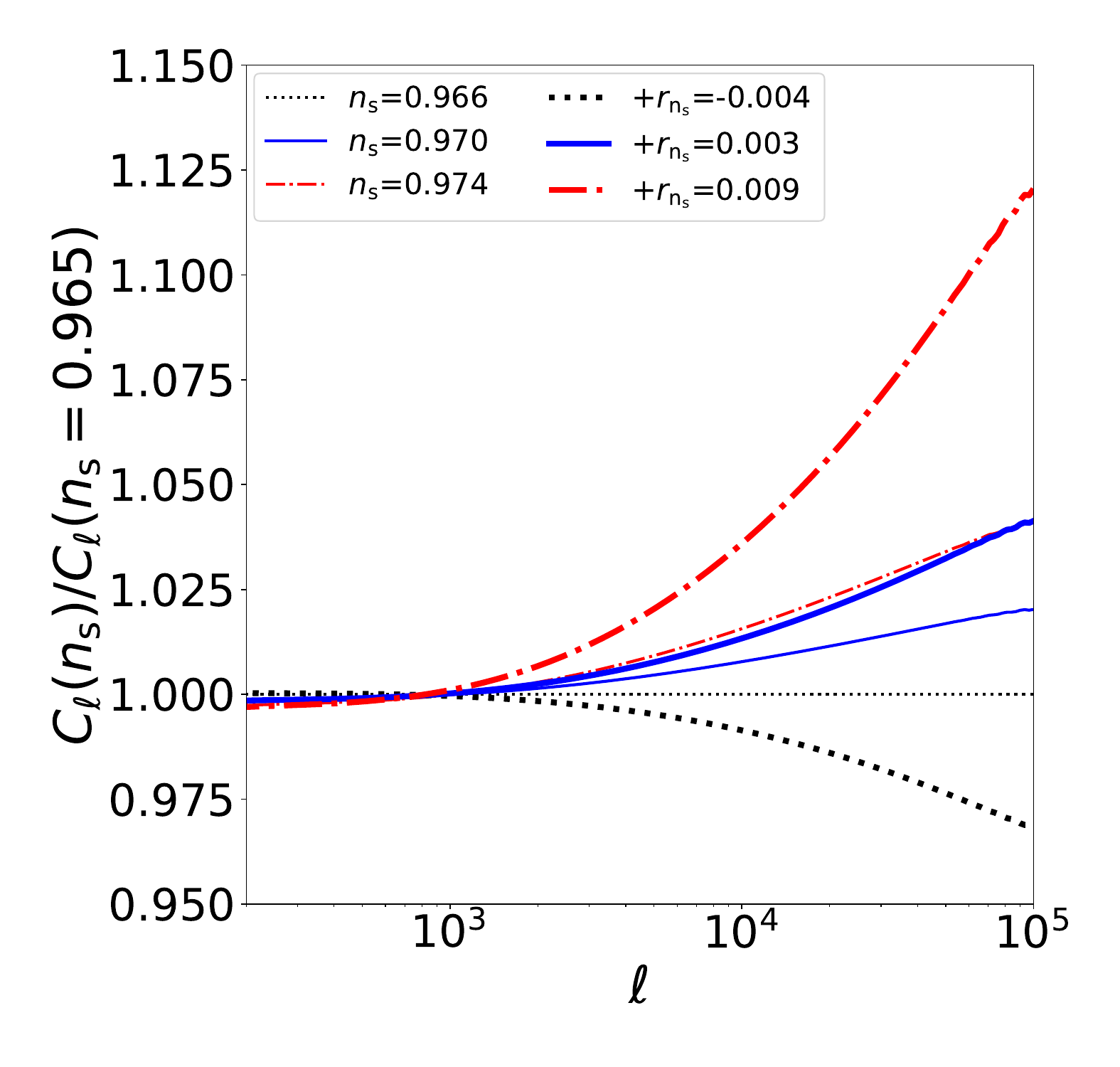}
    \caption{The difference of anisotropies of CMB temperature induced by Compton scattering with the six parameter sets summarized in Table.~\ref{table: ns_rns}. We normalized them by the parameter set (I), $(n_{\m{s}},r_{n_{\m{s}}})=(0.9659,0)$.}
    \label{fig: Cell_sz_ratio_ns_run_pl}
\end{figure}

Next, we consider the thermal SZ effect. 
The thermal SZ effect is the inverse Compton scattering of CMB photons caused by hot electrons inside the dark matter halos.
The brightness temperature of the CMB photons is lifted up by this thermal SZ effect and, as a result, CMB temperature anisotropy is produced, following the distribution of dark matter halos in the sky.
Therefore, the SZ effect temperature anisotropy is sensitive to the dark matter halo abundance.
Using the same gas model as explained in Sec.~\ref{sec: bgff_from_lcdmhalos}, we calculate the thermal SZ effect temperature anisotropy and investigate the signal dependence on the six parameter sets.
In Fig.~\ref{fig: Cell_sz_ratio_ns_run_pl}, we plot them as the ratio of the thermal SZ signals like in Fig.~\ref{fig: Cell_ratio_ns_run_pl}.

In the case of the thermal SZ effect, the angular power spectrum is sensitive to the abundance of DM halos with mass $M_{\m{halo}}\gtrsim 10^{13}M_{\odot}$ as shown in Ref.~\cite{2002MNRAS.336.1256K}.
The mass variance of such DM halos is not sensitive to the spectral index and the running because the scale corresponding to these mass scales is close to the pivot scale in Eq.~\eqref{eq: power_def}. 
Therefore, the dependence of the signal on the parameter set is not strong.
However, in smaller scales~($\ell\gtrsim 10^3$), the 
inner structure of such large DM halos affects the thermal SZ anisotropy.
The concentration parameter which determines the inner structure depends on the spectral index and running in our model. 
Therefore, the dependence on the parameter set becomes stronger.
However, even so, the difference is only by $\sim 15\%$ at best in the very small scale, $\ell\sim10^5$.

Compared with these CMB observations accessing the primordial curvature perturbations,
the free-free signals have a stronger dependence on the spectral index and the running,
because the most contribution of the free-free signal comes from DM halos with smaller mass which other observations are not sensitive to. 
Consequently, these results suggest that the observation of free-free emission could be a powerful probe of the small-scale primordial perturbations, in particular, the spectral index and the running.

% \subsubsection{Compared to the tSZ cosmology}
% To represent the extent of the dependency, we also show the anisotropy induced by the thermal Sunyaev-Zel'dovich effect of halos for the the comparison.
% In the case of tSZ effect, the angular power spcetrum is sensitive to the number of the high mass halo, $M_{\m{halo}}\gtrsim 10^{13}M_{\odot}$ as shown in Ref.~\cite{2002MNRAS.336.1256K}.
% The mass variance of  such high mass is not sensitive to the spectral index because these corresponding scales are near the pivot scale for the spectral index.

% \begin{figure}[htp]
%     \centering
%     \includegraphics[width=1.0\hsize]{images/D_ell_sz_lambdaCDM_jeans_mass_ns_run_planck.pdf}
%     \caption{Anisotropy of the CMB temperature induced by Compton scattering through halos. The line color and style mean the each model summarized in Table.~\ref{table: ns_rns}.}
%     \label{fig: Dell_ns_sz}
% \end{figure}

\section{Conclusion}\label{sec: conclusion}

In this paper, we have investigated cosmological free-free emission originating dark matter halos in the $\Lambda$CDM model.
%The free-free emission depends on the gas profile in DM halos.
%Adopting the analytic gas model which follows the hydrostatic equilibrium in the DM NFW profile,
%we have evaluated the free-free emission from one individual halo.
We adopt the analytic gas model which follows the hydrostatic equilibrium in the DM NFW profile to estimate the free-free emission from individual halos.
Using this result, we have calculated the global signal and the anisotropy,
considering the cosmological structure formation based on the Press-Schechter formalism.

The evaluated amplitude of the global free-free emission spectrum is roughly $10~\rm Jy/str$
and is nearly frequency independent in the CMB and radio frequency range.
With regard to the anisotropy, the Poisson contribution basically dominates, and the resultant angular power spectrum is proportional to the square of the multipole.
However, the clustering contribution would be comparable to or larger than the Poisson one on large scales, $\ell < 1000$.
Therefore, the anisotropy of cosmological free-free emission on large scales
traces matter density fluctuations.
We have also done the above calculations for the constant gas profile model to confirm the dependence on the gas model.
This constant gas profile model could provide the lowest signal. We have found that the amplitude of the free-free signal becomes $5~\rm Jy/str$.

In addition, we have investigated the mass distribution of the free-free signals. 
We have shown that most of the contribution comes from DM halos with mass $\sim 10^{11} M_\odot$ for
both the global signal and the anisotropy.
This tells us that the measurement of cosmological free-free signals has the potential to probe 
the abundance of such small-mass DM halos, that is, the primordial curvature perturbations on small scales which is the origin of these halos.

As an application of probing the small-scale perturbations in the cosmological context, we examine the sensitivity of the cosmological free-free signal on the spectral index and the running.
The difference in these parameters modifies the amplitude of the global signal and
creates the deviation from the scale dependence with $\propto \ell^2$ in the angular power spectrum.
Our result shows the free-free signals are altered by $20~\%$ even in the parameter set which is consistent with the Planck result.
This modification is larger than the one of the CMB temperature anisotropy induced by the halo thermal SZ effect, which changes the signal by $15~\%$ for the same parameter set.
Therefore, the measurement of the cosmological free-free signals could provide a more stringent constraint on the spectral index and the running.

However, the measurement of the cosmological free-free signal is challenging. 
Although the free-free emission in the sky has been investigated in the CMB and radio frequency range, the observed free-free emission sky is dominated by the signals from the Milky Way Galaxy. 
They are roughly 100 times larger in the all-sky average and 10 times larger in the high galactic latitude than the cosmological free-free emission obtained in this paper.
Therefore, the cosmological free-free signal cannot be measured without removing the Milky Way contribution with high accuracy.
The future radio observation, Square Kilometre Array, has enough sensitivity to measure the cosmological free-free signals.
The cross-correlation study is useful to reduce foreground contamination.
The clustering effect of DM halos due to the underlying matter density fluctuations
can make a non-negligible contribution to the cosmological free-free emission anisotropy on large scale. 
Therefore, the cross-correlation with other cosmological probes of the matter density fluctuations
including galaxy survey or future 21-cm intensity map around $z\sim 1$ could be useful to reveal the cosmological free-free anisotropy on large scales.
We have addressed this issue in the next paper.

The theoretical uncertainty of the gas profile in DM halos is also one of the difficulties
to probe the dark matter halo abundance and the primordial curvature perturbations.
The calibration by other cosmological observations, e.g., x-ray and SZ cluster observations
might be useful, although 
the free-free emission signals come from smaller DM halos than in other observations.
In this study, we show that the difference of the gas profile model arises from the overall amplitude of the anisotropy, and they do not modify its scale dependence, $D_\ell \propto \ell^2$.
On the other hand, a scale dependence of the primordial curvature perturbations, the spectral index, and the running, change the scale dependence.
Therefore, the detailed measurement of the scale dependence in the anisotropy
can help to solve the degeneracy of the uncertainties between 
the gas profile model and the scale dependence of the primordial curvature perturbations.

\acknowledgements
This work is supported by the Japan Society for the Promotion of Science~(JSPS) KAKENHI Grants No.~JP20J22260 (K.T.A.),
No.~JP21K03533, and No.~JP21H05459 (H.T.).

\bibliography{article}

\end{document}